# AMRec: An Intelligent System for Academic Method Recommendation

**Shanshan Huang, Xiaojun Wan and Xuewei Tang**

Institute of Computer Science and Technology, Peking University, Beijing 100871, China
{huangshanshan2010, wanxiaojun, tangxuewei}@pku.edu.cn

## Abstract

Finding new academic Methods for research problems is the key task in a researcher's research career. It is usually very difficult for new researchers to find good Methods for their research problems since they lack of research experiences. In order to help researchers carry out their researches in a more convenient way, we describe a novel recommendation system called AMRec to recommend new academic Methods for research problems in this paper. Our proposed system first extracts academic concepts (*Task*s and *Method*s) and their relations from academic literatures, and then leverages the regularized matrix factorization Method for academic Method recommendation. Preliminary evaluation results verify the effectiveness of our proposed system.

## Motivation

For researchers in the computer science area, finding new academic *Method*s (e.g. "*graph-based ranking algorithm*") for research problems or tasks (e.g. "*document summarization*") is the key issue during their research career. Researchers should investigate in a field by reading lots of academic literatures first, and then propose their own ideas through thinking, analysis, and repeated experimental trials. This issue is more severe for new researchers, and they need to spend much time reading and learning before they could have thought out some new academic *Method*s for the research problems they are interested in. Hence we think about that if machine can automatically recommend new academic *Method*s for research problems or tasks as references, the research burden for researchers will be largely alleviated, and the research productivity will be much improved to some extent. In recent years, recommendation systems and techniques have been widely investigated in research communities of information retrieval, machine learning, and data mining (Gori et al., 2006; Linden et al, 2003; Chandrasekaran et al., 2008; Tang, 2012). However, these systems and techniques mainly focus on recommendation of documents, products or friends, and the challenging task of academic *Method* recommendation has not yet been attempted in previous work.

We acknowledge that finding new *Method*s is a very difficult task, and many new *Method*s are proposed based on researchers' talent, and this kind of *Method*s can be hardly recommended by machine. However, a number of *Method*s can be acquired based on the similarity or analogy of *Method*s or *Tasks*, which makes automatic *Method* recommendation possible. For example, if two *Task*s (e.g. "*document summarization*" and "*keyphrase extraction*") share common characteristics, a *Method* (e.g. "*graph-based ranking algorithm*") having been adopted for one *Task* may be suitable for the other *Task*, too. On the other hand, if two *Method*s share common characteristics, and one *Method* has been applied to a *Task*, then the other *Method* may be suitable for the same *Task*. In practice, researchers usually read papers on other related research problems to find new *Method*s for their own research problems. Note that we do not aim to recommend totally new *Method*s which have not been applied to any *Task* but recommend for a specific *Task* with new *Method*s which have been applied to other *Task*s.

## System Description

### Framework

The framework of our AMRec system is shown in Figure 1, which mainly consists of two modules: concept and relation extraction, and academic *Method* recommendation. The extraction module aims to extract the *Task* and *Method* concepts and their relations from academic articles. The recommendation module aims to recommend new *Method*s for specific *Task*s by utilizing the existing relations between the two kinds of concepts. The recommendation process is actually a link prediction process, which predicts new links between the *Task* concepts and the *Method* concepts. The two modules will be introduced in the next two subsections, respectively.

### Concept and Relation Extraction

Two kinds of concepts are firstly extracted by using CRF (Lafferty et al., 2001) and a few hand-crafted rules: *Task* concepts and *Method* concepts. *Task* concepts are specific problems to be solved in academic literatures, including all concepts related to tasks, subtasks, problems and projects, like "machine translation", "document summarization", etc. *Method* concepts are defined as ways to solve specific *Task*s, including all concepts describing algorithms, techniques, models, tools and so on, such as "Markov logic", "CRFs", and "heuristic-based algorithm". The features used in CRF include word-based, POS-based and

keyword-based features. The rules are used for guaranteeing the extraction precision.

The relations between concepts are then extracted by using SVM (Cortes and Vapnik, 1995) and a few hand-crafted rules. Each pair of concepts is classified to determine whether there exists a relation between them. The features include phrase length and position information, relation-related keywords and their position information. Finally, the *Task-Method*, *Task-Task* and *Method-Method* relations are extracted, and a few rules are used for improving the extraction precision.

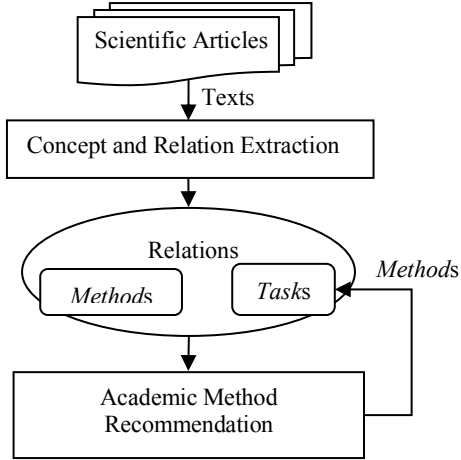

Figure 1: System framework

## Academic Method Recommendation

Matrix factorization (MF) (Ma and Hao, 2008) is one of the most popular recommendation models, and our recommendation approach incorporates the relations within the same kind of concepts into matrix factorization models to improve the recommendation performance. Since matrix factorization models map both *Task* and *Method* concepts to a unified latent factor space, related concepts should have similar latent factor vectors. We add this constraint into matrix factorization by adding one of the following concept relation regularization terms to the original terms:

$$\min_{T,M} L(R,T,M) = \frac{1}{2}\sum_{i=1}^{m}\sum_{j=1}^{n} C_{ij}(R_{ij} - T_i^T M_j)^2 + \frac{\lambda_t}{2}\|T_i\|_F^2$$
$$+ \frac{\lambda_m}{2}\|M_j\|_F^2 + \frac{\beta}{2}xRR,$$

$$xRR = \begin{cases} \sum_{i=1}^{m}\left\|T_i - \sum_{k\in\mathbb{C}_T^-(i)} weig(i,k)T_k\right\|_F^2 & for\ MF-TRR \\ \sum_{i=1}^{n}\left\|M_i - \sum_{k\in\mathbb{C}_M^-(i)} weig(i,k)M_k\right\|_F^2 & for\ MF-MRR \end{cases}$$

where $T_i$ and $M_i$ denote the latent factor vectors for each *Task* and each *Method*, respectively. $R_{ij}$ is the relation matrix to describe relations between *Tasks* and *Methods*. The rows in matrix denote *Tasks*, and the columns represent *Methods*. $Weig(i,k)$ is the function for measuring the strength of concept relations by using SimRank (Jeh et al., 2002), $\mathbb{C}_T^-(i)$ is the set of *Tasks* that $T_i$ has relation with (e.g. evolved from), and $\mathbb{C}_M^-(i)$ is the set of *Methods* that $M_i$ has relation with (e.g. enhanced on, evolved from or based on). Usually, a parameter $\beta$ is used to make trade-off between the original terms and the newly added regularization term.

For model learning, the alternating least squares (ALS) algorithm (Lin and Chih-Jen, 2007) is adopted. After we obtain the latent factor vectors for *Tasks* and *Methods*, *Methods* are recommended for a *Task* based on the inner products of the latent factor vectors between *Methods* and the *Task*. For each *Task*, we recommend a ranked list of *Methods* with high results.

## Evaluation and Discussion

Preliminary evaluation is conducted on a dataset consisting of 9754 research articles in the NLP field. Articles published before 2008 (including 2008) construct the training data set and the other articles published after 2009 (including 2009) construct the testing data set. We extract 9863 *Methods* and 3862 *Tasks* for the training data, 6411 *Methods* and 2792 *Tasks* for the test data, 12862 relation pairs for the training data and 10339 relation pairs for the test data. We exclude the *Task-Method* relations from the test data if such relations have already appeared in the training dataset, and the recommendation results of different algorithms are evaluated based on the new *Task-Method* relations in the test dataset. The average precision values (P) at top *N* are used as evaluation metrics, indicating the ratio of *Methods* recommended correctly to *Tasks*. Our proposed models (MF-TRR and MF-MRR) are compared with the MF and traditional collaborative filtering (CF) models in Table 1.

Table 1: Comparison results (%)

| Method | P@10 | P@30 | P@50 |
|---|---|---|---|
| **CF** | 4.56 | 4.15 | 3.95 |
| **MF** | 5.98 | 5.31 | 4.98 |
| **MF-TRR** | 7.57 | 7.39 | **7.36** |
| **MF-MRR** | **7.64** | **7.43** | 7.35 |

We can see that our proposed models can outperform the baseline models over all metrics, which means that the concept relation regularization terms can improve the performance. We also find that our proposed models can perform better than the baselines when $\beta$ is set in a wide range of values.

However, the *Method* recommendation performance is not high, which means that it is actually a very difficult task and needs more investigation in the future. We will also download more literature articles to improve the coverage of concepts and relations, which will make the results more reliable.